\documentstyle[preprint,tighten,aps,floats,psfig,amssymb]{revtex}

\begin{document}
\draft
\title{
Critical structure factor in Ising systems
}
\author{Victor Mart\'\i n-Mayor,$^{1,2}$ Andrea Pelissetto,$^1$ 
and Ettore Vicari$^3$ }
\address{$^1$
Dipartimento di Fisica dell'Universit\`a di Roma La Sapienza
and INFN, I-00185 Roma, Italy 
}
\address{$^2$
Statistical Mechanics Center (SMC), INFM, I-00185 Roma, Italy 
}
\address{$^3$ 
Dipartimento di Fisica dell'Universit\`a di Pisa
and INFN, I-56127 Pisa, Italy
\\
{\bf e-mail: \rm
{\tt Victor.Martin@roma1.infn.it},
{\tt Andrea.Pelissetto@roma1.infn.it},
{\tt Vicari@df.unipi.it}
}}

\date{\today}
\maketitle

\begin{abstract}
We perform a large-scale Monte Carlo simulation of the three-dimensional
Ising model on simple cubic lattices of size $L^3$ with $L=128$ and 256.
We determine the corresponding structure factor (Fourier transform of the 
two-point function) and compare it with several approximations 
and with experimental results. We also compute the turbidity as a function 
of the momentum of the incoming radiation, focusing in particular 
on the deviations from the Ornstein-Zernicke expression of 
Puglielli and Ford.
\end{abstract}




\section{Introduction}

Near a phase-transition critical point, some observed quantities show
a universal behavior that is common to a large class of systems,
independently of the microscopic details. A very important 
universality class is the Ising one that is characterized by
short-range interactions and a scalar order parameter.
It describes the liquid-vapor transition in fluids, the mixing 
transition in multicomponent systems, the Curie transition 
in (anti)ferromagnets with axial anisotropy. The Ising critical
behavior has been extensively studied both theoretically and 
experimentally, see Refs. \cite{Cargese-review,PV-review}. In 
particular, the critical exponents, the equation of state, and several
amplitude ratios have been determined with good precision. 
Another important quantity in the theory of critical 
phenomena is the static structure factor, that can be 
measured experimentally by determining the intensity of 
the light scattered by the fluid relative to the intensity of the 
incident light \cite{foot1}. To probe larger wave numbers, neutrons are used 
instead of light. At the critical density of fluids near the 
gas-liquid critical point or at the critical concentration of binary 
fluids near the critical mixing point, one expects for 
$t\equiv (T - T_c)/T_c \to 0$ the general scaling 
behavior\cite{Fisher-64,FB-67,FA-74}
\begin{equation}
S_\pm(k) = \chi g_\pm (k\xi),
\label{scaling-Sk}
\end{equation}
where $\chi = C^\pm |t|^{-\gamma}$, $\xi$ is the correlation length
which diverges as $\xi = f^\pm |t|^{-\nu}$, $k$ is the 
momentum-transfer vector, and $\pm$ refers to the two phases, 
$+$ (resp. $-$) corresponding to the high- (resp. low-) temperature phase. 
Its absolute value is given by
\begin{equation} 
k = {4 \pi\over \lambda} \sin {\theta\over2},
\end{equation}
where $\lambda$ is the wavelength of the radiation (neutrons) in the 
scattering medium and $\theta$ is the scattering angle. The functions 
$g_\pm(Q)$, normalized so that 
\begin{equation} 
g^{-1}_\pm(Q) = 1 + Q^2 + O(Q^4)
\end{equation}
for $Q\equiv k\xi\to 0$ 
(this defines $\xi$ as the second-moment correlation length), 
are universal. Their limiting behavior is well known. 
For $Q$ small, $g_\pm(Q)$ is approximated by the leading term, the 
so-called Ornstein-Zernicke approximation
\begin{equation} 
g_{OZ}(Q) = {1\over 1 + Q^2}.
\end{equation}
Such an approximation well describes the data up to $Q\approx 1$ and 
is routinely used in the analysis of the data with $k\xi$ small 
and of the turbidity for the determination of the correlation
length \cite{PF-70}. On the other hand, for large $Q$, 
$g_\pm(Q)$ shows an anomalous decay controlled by the exponent $\eta$ 
\begin{equation}
g_\pm(Q) \approx {C_1^\pm\over Q^{2-\eta}}.
\end{equation}
Therefore, the experimental determination of the structure factor
for large wavenumbers allows a direct determination of the 
exponent $\eta$ 
\cite{BBC-82,CBS-79,SBSWC-80,DLC-89,Izumi-89,JSS_92,SKK-96,LMBWH-97,%
BBB-97,BC-97-98,DLMFL-98,BRCB-00,BCB-01}.

In this paper, we compute the structure factor in the high-temperature phase
for small values of $Q$ by means of Monte Carlo simulations on 
lattices $L^3$, with $L=128, 256$. We are able to determine the function 
$g_+(Q)$ with an error of less than 1\% (resp. 2\%) for $Q\lesssim 5$
(resp. $Q\lesssim 20$). These numerical results together with the 
most recent estimates of the critical exponents \cite{CPRV-02}
are then used to determine interpolations that are valid for all values of $Q$
and have the correct large-$Q$ behavior. For this purpose, we use
a dispersive approach \cite{FS-75,Bray-76,FB-79}, which allows us to 
determine an interpolating form for $g_+(Q)$ that agrees with the Monte Carlo
data in the small-$Q$ region and that well approximates (within 0.5\%)
the experimental results of Ref. \cite{DLMFL-98}.

These results are then used to compute the turbidity, i.e. the 
attenuation of the transmitted light intensity per unit optical path length
due to the scattering with the sample. This quantity is routinely measured in
experiments, since it allows the determination of the correlation 
length. In particular, we compute the deviations from the 
Puglielli-Ford expression \cite{PF-70}, that is based on the 
Ornstein-Zernicke approximation.

The paper is organized as follows.  
In Sec. \ref{sec.2} we review the theoretical results for the structure
factor. In Sec. \ref{sec.2.1} we define the basic observables and 
report the behavior of $g_\pm(Q)$ for small and large values of $Q$. 
Estimates of the constants appearing in these expansions are reported 
in Sec. \ref{sec.2.2}. In Sec. \ref{sec.2.3} we discuss Bray's approximation.
First, we discuss the high-temperature phase: we update the estimates of 
Ref. \cite{Bray-76} by using the most recent results for the critical 
exponents. Then, we generalize the approximation to the low-temperature 
phase.
In Sec. \ref{sec.3} we discuss our high-temperature 
Monte Carlo results which are 
compared with approximate expressions and with the experimental data
of Ref. \cite{DLMFL-98}. In Sec. \ref{sec.4} we compute the turbidity,
focusing on the deviations from the Puglielli-Ford expression \cite{PF-70} 
due to the anomalous decay of $g_+(Q)$. 
We find that the turbidity is larger than this expression by
1\% (resp. 5\%) for $Q_0 = 15$ (resp. 350), where $Q_0 = q_0 \xi$ and 
$q_0$ is the momentum of the incoming radiation.

\section{Theoretical results} \label{sec.2}

\subsection{Definitions} \label{sec.2.1}

Several theoretical results are available for the structure factor.
For $Q$ small, one can compute the corrections to the Ornstein-Zernicke
behavior, by writing 
\begin{equation} 
g^{-1}_\pm (Q) = 1 + Q^2 + \sum_{n=2} c_n^\pm Q^{2n}.
\label{gQ-small}
\end{equation}
For large $Q$, the structure factor behaves as 
\begin{equation}
g_\pm (Q) \approx {C_1^\pm\over Q^{2 - \eta}}
  \left(1 + {C_2^\pm\over Q^{(1-\alpha)/\nu}} +
            {C_3^\pm\over Q^{1/\nu}}\right),
\label{eq:FL}
\end{equation}
a behavior predicted theoretically by Fisher and Langer 
\cite{FL-68} and proved in the field-theoretical 
framework in Refs. \cite{BAZ-74,BLZ-74}. 

\begin{table}[tbp]
\caption{
Estimates of $c_{n}^\pm$, $S_M^\pm$, and $S_Z^\pm$. 
IHT denotes the results obtained from
the analysis of high-temperature expansions for 
improved models, HT,LT results obtained from
the analysis of high- and low-temperature expansions for the 
Ising model, while 
``$\epsilon$-exp." and ``$d=3$ $g$-exp." label the field-theoretical 
results.
(sc) and (bcc)
denote the simple cubic and the body-centered cubic lattice respectively.
Unless stated otherwise, field-theoretical results are taken from 
Ref.~\protect\cite{CPRV-99}, 
while the IHT estimates are taken from Ref.~\protect\cite{CPRV-02}.
For $S_M^-$ we should also report the Monte Carlo estimate of 
Ref.\protect\cite{ACCH-97}, $S_M^- = 0.941(11)$.
}
\label{cn-table}
\begin{tabular}{cr@{}lr@{}lr@{}lr@{}l}
\multicolumn{1}{c}{}&
\multicolumn{2}{c}{IHT}&
\multicolumn{2}{c}{HT,LT}&
\multicolumn{2}{c}{$\epsilon$-exp.}&
\multicolumn{2}{c}{$d=3$ $g$-exp.}\\
\tableline \hline
$c_2^+$ &
$-$3&.90(6)$\times 10^{-4}$ & $-$3&.0(2)$\times 10^{-4}$ \cite{CPRV-98}&
$-$3&.3(2) $\times 10^{-4}$ & $-$4&.0(5) $\times 10^{-4}$  \\
& && $-$5&.5(1.5)$\times 10^{-4}$ (sc) \cite{TF-75} & && &\\
& && $-$7&.1(1.5)$\times 10^{-4}$ (bcc) \cite{TF-75} & && & \\\hline
$c_3^+$ &
0&.88(1)$\times 10^{-5}$ & 1&.0(1)$\times 10^{-5}$ \cite{CPRV-98} & 0&.7(1) $\times 10^{-5}$ &
1&.3(3) $\times 10^{-5}$ \\
& && 0&.5(2)$\times 10^{-5}$ (sc) \cite{TF-75} & && & \\    
& && 0&.9(3)$\times 10^{-5}$ (bcc) \cite{TF-75} & && &  \\\hline
$c_4^+$ & $-$0.&4(1)$\times 10^{-6}$  & & &
$-$0&.3(1)$\times 10^{-6}$ & $-$0&.6(2)$\times 10^{-6}$\\ \hline
$S_M^+$ & 0&.999601(6) &   0&.99975(10) \cite{CPRV-98} & 0&.99968(4) & 
0&.99959(6)  \\ \hline
$S_Z^+$ & 1&.000810(13) &    & & & & & \\ \hline\hline
$c_2^-$ &
&& $-$1&.2(6)$\times 10^{-2}$ \cite{TF-75} &
$-$2&.4$\times 10^{-2}$ \cite{CDK-75} & & \\ \hline
$c_3^-$ &
&& 7&(3)$\times 10^{-3}$ \cite{TF-75} &
3&.9$\times 10^{-3}$ \cite{CDK-75} & & \\ \hline
$S_M^-$ &
&& 0&.938(8) \cite{CPRV-99} && && \\
& && 0&.930(6) \cite{FZ-98} && && \\
\end{tabular}
\end{table}

Beside the constants $c_n^\pm$, the constants 
$S_M^\pm$ and $S_Z^\pm$ defined by
\begin{eqnarray}
S_M^\pm&\equiv&M_{\rm gap}^2 \xi^2,\label{SMdef}\\
S_Z^\pm&\equiv& \chi/({\xi^2 Z_{\rm gap}}),\label{SZdef}
\end{eqnarray}
are of theoretical interest.
Here $M_{\rm gap}$ (the mass gap of the theory) and $Z_{\rm gap}$
determine the long-distance behavior of the two-point function in 
$x$-space:
\begin{equation}
G(x)\approx  {Z_{\rm gap}\over 4\pi |x|} e^{-M_{\rm gap}|x|}.
\label{largexbehavior}
\end{equation}
The critical limits of $S_M^\pm$ and $S_Z^\pm$ are related to the 
imaginary zeroes
$\pm i Q_0$ of $g^{-1}_\pm(Q)$ closest to the origin by
\begin{eqnarray}
S_M^\pm&=&-Q_0^2,\\
S_Z^\pm&=& \left. {dg^{-1}(Q) \over dQ^2} \right|_{Q=\pm i Q_0}.
\end{eqnarray} 

\subsection{Numerical results} \label{sec.2.2}

The coefficients $c_n^+$ turn out to be very small \cite{FA-74}, 
$c_2^+\sim 10^{-4}$, 
and this explains the success of the Ornstein-Zernicke approximation 
up to $Q\sim 1$. The constants $c_n^+$ have been calculated by
field-theoretic methods.  They have been computed to $O(\epsilon^3)$
in the framework of the $\epsilon$-expansion \cite{Bray-76}, and to
$O(g^4)$ in the framework of the $d$=3 $g$-expansion
\cite{CPRV-98}. The perturbative series have been resummed in 
Ref. \cite{CPRV-99} obtaining the results reported in Table 
\ref{cn-table}. The most precise estimates have been obtained 
from the analysis of their high-temperature expansions in improved 
models \cite{CPRV-02}, see the results labelled by IHT in 
Table \ref{cn-table}.

As already observed in Ref.\ \cite{CPRV-98}, the coefficients show
the pattern
\begin{equation}
|c_n^+|\ll |c_{n-1}^+|\ll...\ll |c_2^+| \ll 1\qquad\qquad 
{\rm for}\qquad n\geq 3.
\label{patternci}
\end{equation}
Therefore, a few terms of the expansion of $g_+(Q)$ in powers of $Q^2$
provide a good approximation of $g_+(Q)$ in a relatively large region around
$Q=0$: as we shall see, deviations are less than 1\% up to 
$Q\approx 3$. This is in agreement with the
theoretical expectation that the singularity of $g_+(Q)$ nearest to the
origin is the three-particle cut \cite{FS-75,Bray-76}.
If this is the case, the convergence radius $r_{+}$ of the Taylor
expansion of $g^{-1}_+(Q)$ is $r_+=3\sqrt{S_M^+}$.  Since, 
see Table \ref{cn-table},
$S_M^+\approx 1$, at least asymptotically we should have
\begin{equation}
c_{n+1}^+\approx - {1\over 9}c_n^+.
\label{pattern-cip1-ci}
\end{equation}
This behavior can be checked explicitly in the large-$N$ limit of the
$N$-vector model \cite{CPRV-98}. 

The coefficients $c_n^-$ are also quite small, although not as much as in 
the high-temperature case. Indeed, $c_2^- \approx 10^{-2}$, see Table
\ref{cn-table}. They have been computed using field-theoretical methods
\cite{CDK-75} and from the analysis of low-temperature series 
\cite{TF-75}. In the low-temperature phase, one also observes the 
pattern (\ref{patternci}), although the coefficients decrease slower. 
This is related to the fact that in the low-temperature phase
the nearest singularity is the two-particle cut, so that 
convergence radius $r_-$ of the Taylor
expansion of $g^{-1}_-(Q)$ is $r_-=2\sqrt{S_M^-}$, and therefore,
\begin{equation}
c_{n+1}^- \approx - {1\over 4 S_M^-} c_n^- \approx 
   - 0.27 c_n^-.
\end{equation}

The large-order coefficients 
$C_1^\pm$, $C_2^\pm$, and $C_3^\pm$ have been computed 
theoretically within the $\epsilon$ expansion to order $\epsilon^3$
\cite{Bray-76} in the high-temperature phase and to order 
$\epsilon^2$ in the low-temperature phase \cite{CDK-75}.
Using the $\epsilon$-expansion results, we obtain
\begin{equation}
C_1^+ \approx 0.92, \qquad C_2^+ \approx 1.8, \qquad C_3^+ \approx - 2.7.
\label{FL-par-eps}
\end{equation} 
The corresponding low-temperature parameters $C_n^-$ can be derived from 
the high-temperature $C_n^-$ by using a set of relations derived in 
Ref. \cite{BLZ-74}: 
\begin{eqnarray}
{C_1^+\over C_1^-} &=& U_2^{-1} U_\xi^{2 - \eta} \nonumber \\
{C_2^+\over C_2^-} &=& - U_0 U_\xi^{(1-\alpha)/\nu} \nonumber \\
{C_3^+\over C_3^-} &=& - U_\xi^{1/\nu} ,
\label{ratioCpmn}
\end{eqnarray}
where
\begin{eqnarray}
&& U_0 = {A^+\over A^-}, \nonumber \\
&& U_2 = {C^+\over C^-}, \nonumber \\
&& U_\xi = {f^+\over f^-}. 
\end{eqnarray}
Here, $C^\pm$ and $f^\pm$ are the amplitudes of the susceptibility and of the 
second-moment correlation length defined above, while $A^\pm$ are defined from
the critical behavior of the specific heat, $C_H \approx A^\pm |t|^{-\alpha}$.
Using the estimates of Ref. \cite{CPRV-02}, we obtain
\begin{eqnarray} 
C_1^- &=& 1.275(10)\,  C_1^+ \hphantom{,} \approx 1.17, \nonumber \\
C_2^- &=& -0.728(5)\,  C_2^+ \approx -1.3, \nonumber \\
C_3^- &=& -0.345(2)\,  C_3^+ \approx 0.9 . 
\end{eqnarray}
The large-momentum  behavior of the structure factor has also been studied 
experimentally and the behavior (\ref{eq:FL}) has been explicitly 
verified in the high-temperature phase. 
In particular, the exponent $\eta$ and the constant $C_1^+$ 
have been determined. 
Analysis of the large-$k$ 
behavior of the structure factor $S_+(k)$ gives: 
$\eta = 0.017(15)$, $C_1^+= 0.96(4)$ and 
$\eta \approx 0.030(25)$, $C_1^+\approx 0.95(4)$ (two different 
parametrizations of the structure factor are used)
\cite{CBS-79}; 
$\eta = 0.0300(15)$, $C_1^+\approx 0.92(1)$ \cite{DLC-89};
$\eta = 0.042(6)$, $C_1^+\approx 0.915(21)$ \cite{DLMFL-98}. 
No unbiased determination of $C_2^+$ and $C_3^+$ is available.
Fixing $C_2^+ + C_3^+ = -0.9$ (the $\epsilon$-expansion result
of Ref. \cite{Bray-76}), Ref. \cite{DLMFL-98} 
obtains $C_2^+ = 2.05(80)$ and $C_3^+ = - 2.95(80)$, in reasonable
agreement with the $\epsilon$-expansion predictions.

\subsection{Bray's approximation} \label{sec.2.3}

In order to compare with the experimental data it is important 
to know the function $g_\pm(Q)$ for all values of $Q$. 
For the high-temperature $g_+(Q)$, several interpolations have been proposed
with the correct large- and small-$Q$ behavior 
\cite{FB-67,TF-75,FS-75,Bray-76,FB-79,BBC-82}. 
The most successful one is due to Bray \cite{Bray-76}, 
which incorporates the expected singularity structure of $g_+(Q)$. 
Here, we present Bray's interpolation together with its generalization to the 
low-temperature phase.

In this approach, one 
assumes $g^{-1}_\pm (Q)$ to be well defined in the complex
$Q^2$ plane, with a cut on the negative real $Q^2$ axis,
starting at $Q^2 = - r_\pm^2$, where, as discussed above,
$r_+^2 = 9 S_M^+$, $r_-^2 = 4 S_M^-$ . Then
\begin{eqnarray}
g^{-1}_\pm(Q) &=& {2 \sin \pi\eta/2\over \pi C_1^\pm }
   \int_{r_\pm}^\infty du\, u^{1-\eta} F_\pm(u)\left[{S_M\over u^2 - S_M} + 
       {Q^2 \over u^2 + Q^2}\right],
\label{gQ-dispersive}
\end{eqnarray}
where $F_\pm(u)$ is the spectral function, which must satisfy 
$F_\pm(+\infty) = 1$, $F_\pm(u) = 0$ for $u< r_\pm$, and $F_\pm(u)\ge 0$ for 
$u\ge r_\pm$. 
Notice the appearance of the constant $C_1^\pm$, which is determined,
once $F_\pm(u)$ is given, by requiring $g^{-1}_\pm(0) = 1$. 

In order to obtain an approximation one must specify $F_\pm(u)$. 
Bray \cite{Bray-76} proposed to use a spectral function that gives 
exactly the Fisher-Langer asymptotic behavior, i.e.
\begin{equation}
F_{\pm,B}(u) = {P_\pm(u) - Q_\pm(u) \cot {1\over2} \pi \eta \over 
          P_\pm(u)^2 + Q_\pm(u)^2},
\end{equation}
where 
\begin{eqnarray}
P_\pm(u) &=& 1 + {C_2^\pm\over u^p} \cos {\pi p\over 2} + 
             {C_3^\pm\over u^{1/\nu}} \cos {\pi\over 2\nu}, 
\nonumber \\
Q_\pm(u) &=&  {C_2^\pm\over u^p} \sin {\pi p\over 2} + 
             {C_3^\pm\over u^{1/\nu}} \sin {\pi\over 2\nu}, 
\label{PQ-def}
\end{eqnarray}
with $p\equiv (1-\alpha)/\nu$. These definitions do not specify the 
spectral functions completely since several quantities are 
still unknown. First of all, we should fix the critical exponents.
We will use the estimates of Ref. \cite{CPRV-02}, obtained from the 
analysis of high-temperature expansions for improved models:
\begin{eqnarray}
\gamma &=& 1.2373(2), \nonumber \\
\nu    &=& 0.63012(16), \nonumber \\
\eta   &=& 0.03639(15), \nonumber \\
\alpha &=& 0.1096(5). \label{exponents}
\end{eqnarray}
Several other determinations are reported in Refs. \cite{PV-review,CPRV-99}.
For $S_M^+$ we use the estimate labelled by IHT reported in Table
\ref{cn-table}, while for $S_M^-$ we employ  the low-temperature 
prediction of Ref. \cite{CPRV-99}, see Table \ref{cn-table}. 
We must also fix $C^\pm_2$ and $C^\pm_3$. In the 
high-temperature phase, Bray proposes to fix 
$C_2^++C_3^+$ to its $\epsilon$-expansion value $C_2^+ + C_3^+ = -0.9$ 
and then to determine these constants by requiring $F_{+,B}(r_+) = 0$. 
These conditions completely fix the spectral function and thus 
the structure factor. As a check, we can compare the estimates of 
$c_n^+$ and $C_i^+$ obtained by using Bray's approximation $g_{+,B}(Q)$ 
with the previously quoted results. 
We obtain
\begin{eqnarray}
C_1^+ \approx 0.918, \qquad C_2^+ &\approx& 2.56, \qquad C_3^+ \approx -3.46,
\nonumber \\
c_2^+ \approx - 4.2\cdot 10^{-4}, \quad && \quad  
c_3^+ \approx 1.0\cdot 10^{-5}.
\end{eqnarray}
The constants $C_1^+$, $C_2^+$, and $C_3^+$ are in reasonable 
agreement with the $\epsilon$-expansion results (\ref{FL-par-eps}),
while $c_2^+$ and $c_3^+$ are close to the estimates reported in Table
\ref{cn-table}.

\begin{figure}[tb]
\hspace{-1cm}
\vspace{0cm}
\centerline{\psfig{width=10truecm,angle=-90,file=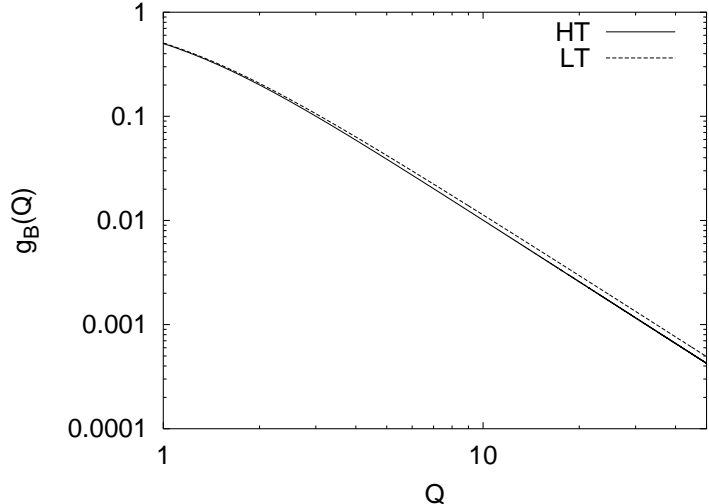}}
\vspace{0cm}
\caption{Scaling functions $g_\pm(Q)$ versus $Q$ in Bray's approximation.
We report the high- (HT) and low- (LT) temperature scaling functions.
}
\label{Fig-Bray}
\end{figure}

In the low-temperature phase, we have tried to follow again Bray's strategy.
We have first set $C_2^- + C_3^- = -0.4$ and required
$F_{-,B}(r_-) = 0$. However, the resulting estimates of $C_n^-$ and $c_n^-$
are not in agreement with the previous results: we find 
$C_1^- \approx 0.87$, $c_2^- \approx - 1\times 10^{-3}$. Little changes 
if we fix $C_2^+ + C_3^+ = -0.4$ and use the relations (\ref{ratioCpmn}).
For this reason, we have given up requiring $F_{-,B}(r_-) = 0$
and we have simply set $C_2^- = -1.3$, $C_3^- = 0.9$, as obtained in
the previous section. Then, Bray's approximation gives
\begin{eqnarray}
C_1^- \approx 1.0, \qquad\qquad
c_2^- \approx -1.1 \times 10^{-2}, \qquad\qquad
c_3^- \approx 1.7 \times 10^{-3},
\end{eqnarray}
which are close to previous estimates. A plot of Bray's approximations 
is given in Fig. \ref{Fig-Bray}. Note that the  
structure factors in the high- and low-temperature phases are very similar.

\section{Monte Carlo results} \label{sec.3}

We determine the structure factor in the region of small $k$---as we shall
see, we are able to reach $k\approx $5-10$/\xi$ by means of a 
large-scale Monte Carlo simulation. We consider the Ising model
on a cubic lattice, i.e. the Hamiltonian 
\begin{equation}
{\cal H} = - \beta \sum_{<i,j>} \sigma_i \sigma_j,
\end{equation}
where $\sigma_i = \pm1$ and the summation is over nearest-neighbor pairs
$<i,j>$. We measure the structure factor
\begin{equation}
S(q;\beta,L) = {1\over3} \sum_{x,y,z} \left( e^{iqx} + e^{iqy} + e^{iqz}\right) 
        \langle \sigma_{(0,0,0)} \sigma_{(x,y,z)} \rangle
\label{def-Gq}
\end{equation}
for three different values of $\beta$ and $L$:
(a) $L = 128$, $\beta = 0.2204$; 
(b) $L = 128$, $\beta = 0.2210$; (c) $L = 256$, $\beta = 0.22145$. 
Of course, in Eq. (\ref{def-Gq}) $q = 2 \pi n/L$, where $n$ is an integer.
In the simulation we used the Swendsen-Wang algorithm, 
starting from random configurations and discarding (2-4)$\times 10^4$ 
iterations. The results of the 
simulations are reported in Table \ref{risultati}. We report the 
number of iterations $N_{\rm it}$, the susceptibility $\chi$,
the second-moment correlation length $\xi$ and $h(q;\beta,L)$,
\begin{equation} 
h(q;\beta,L) \equiv \ln \left[ {(1 + q^2 \xi^2) S(q;\beta,L) \over \chi}
                        \right],
\end{equation} 
which directly measures the deviations from a purely Ornstein-Zernicke
behavior.

\begin{table}
\caption{For the three lattices considered, (a), (b), and (c), we report 
the number of iterations $N_{\rm it}$, the susceptibility $\chi$,
the second-moment correlation length $\xi$ and $h(q;\beta,L)$ 
for $n=q L/(2 \pi)$.}
\label{risultati}
\begin{tabular}{lccc}
 & (a)  & (b) & (c) \\ 
\hline
$N_{\rm it}$ &  $4.35\times 10^6$ & $3.2\times 10^6$ & $2.14 \times 10^6$ \\
$\chi$ & 669.9(4) & 1501(2) & 6339(10) \\
$\xi$  & 13.050(7) & 19.739(14) & 41.16(5) \\
\hline
$n$ & \multicolumn{3}{c}{$h(q;\beta,L)$} \\
1 & $-$0.0009(9) & $-$0.0015(11) & $-$0.0002(17) \\ 
2 & $-$0.0002(11) & 0.0003(14) & 0.0001(25) \\ 
3  & 0.0017(12) & 0.0027(16) & 0.0019(27) \\ 
4  & 0.0039(13) & 0.0065(17) & 0.0042(27) \\ 
5  & 0.0063(13) & 0.0096(18) & 0.0067(28) \\ 
6  & 0.0093(13) & 0.0135(18) & 0.0095(28) \\ 
7  & 0.0128(13) & 0.0179(18) & 0.0123(28) \\ 
8  & 0.0178(13) & 0.0232(19) & 0.0141(28) \\ 
9  & 0.0222(14) & 0.0281(18) & 0.0179(28) \\ 
10 & 0.0270(13) & 0.0335(19) & 0.0204(28) \\ 
11 & 0.0326(14) & 0.0398(18) & 0.0234(29) \\ 
12 & 0.0383(13) & 0.0459(17) & 0.0263(28) \\ 
13 & 0.0438(13) & 0.0521(17) & 0.0290(29) \\ 
14 & 0.0510(13) & 0.0593(18) & 0.0324(29) \\ 
15 & 0.0579(13) & 0.0666(18) & 0.0353(28) \\ 
16 & 0.0647(14) & 0.0736(18) & 0.0380(28) \\ 
17 & 0.0722(13) & 0.0815(18) & 0.0409(29) \\ 
18 & 0.0806(13) & 0.0896(18) & 0.0437(28) \\ 
19 & 0.0887(14) & 0.0986(17) & 0.0478(28) \\ 
20 & 0.0975(13) & 0.1078(18) & 0.0506(29) \\ 
21 & 0.1072(14) & 0.1168(18) & 0.0538(29) \\ 
22 & 0.1158(14) & 0.1271(18) & 0.0576(28) \\ 
23 & 0.1258(14) & 0.1366(18) & 0.0616(28) \\ 
24 & 0.1367(14) & 0.1473(18) & 0.0642(29) \\ 
25 & 0.1472(14) & 0.1583(18) & 0.0676(28) \\ 
\end{tabular}
\end{table}

\begin{figure}[tb]
\hspace{-1cm}
\vspace{0cm}
\centerline{\psfig{width=10truecm,angle=-90,file=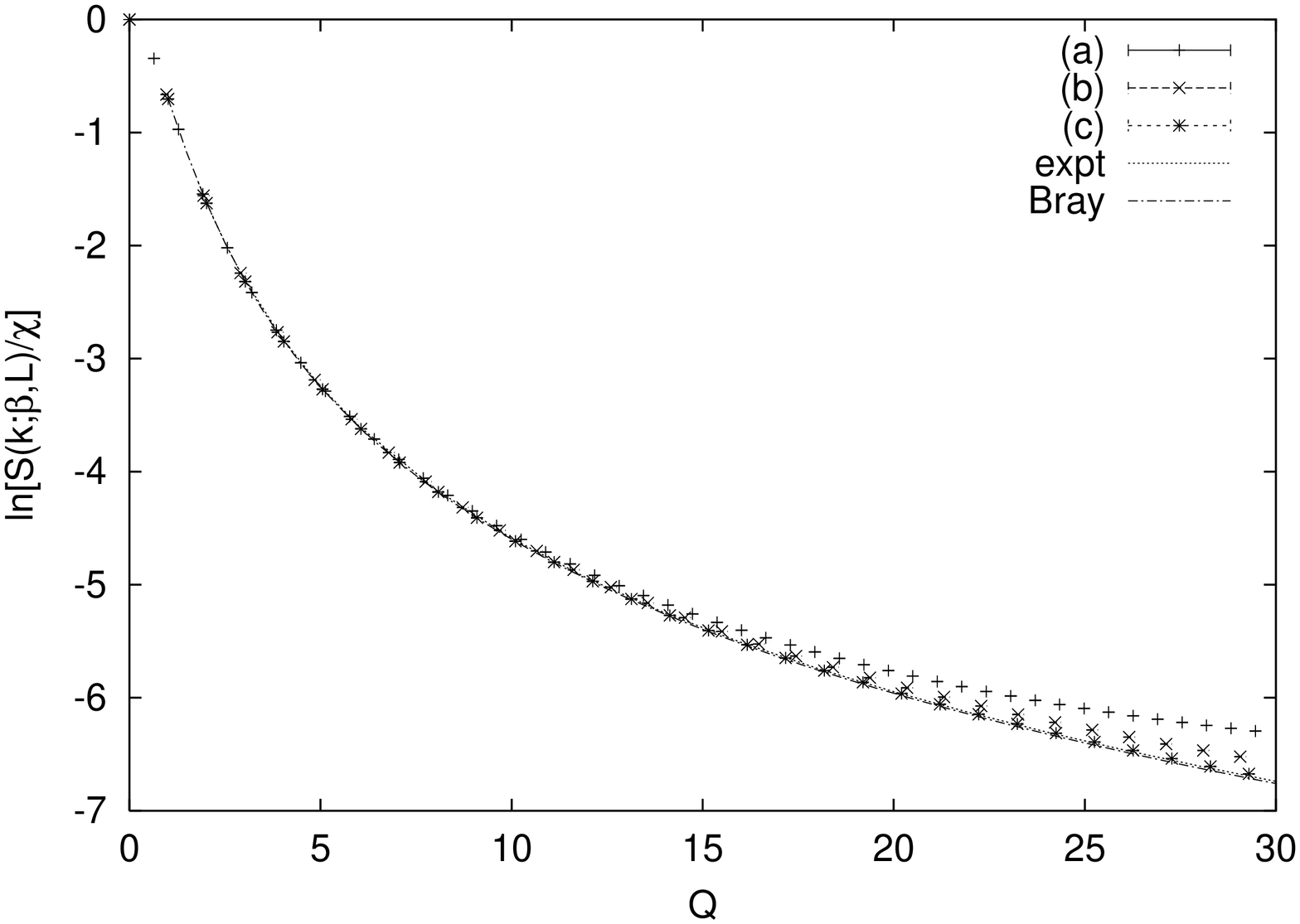}}
\vspace{0cm}
\caption{Function $S(q;\beta,L)/\chi$ versus $Q\equiv q \xi$ for the 
three cases (a), (b), (c). We also report the experimental results 
of Ref.~\protect\cite{DLMFL-98}, ``expt," and Bray's approximation,
``Bray."
}
\label{figgQ}
\end{figure}

In Fig. \ref{figgQ} we plot $S(q;\beta,L)/\chi$ for the three lattices 
considered---errors are smaller than the size of the points---together 
with the experimental results of Ref.~\protect\cite{DLMFL-98} for 
CO$_2$ and 
Bray's approximation. We observe good agreement, the numerical 
data for lattice (c) being close to the experimental ones.

\begin{figure}[tb]
\hspace{-1cm}
\vspace{0cm}
\centerline{\psfig{width=10truecm,angle=-90,file=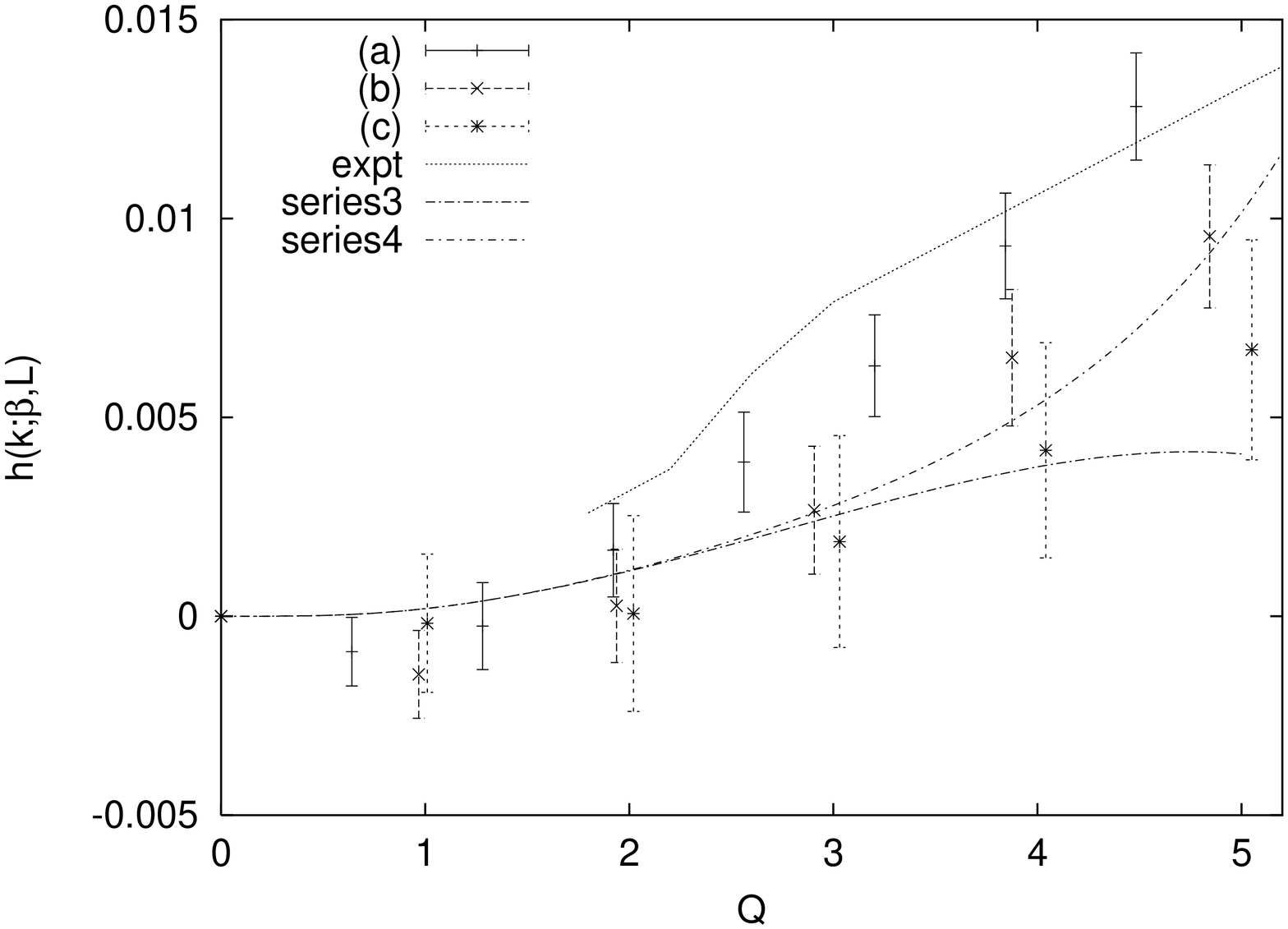}}
\vspace{0cm}
\caption{Function $h(q;\beta,L)$ versus $Q\equiv q \xi$ for the 
three cases (a), (b), (c). We also report the experimental results 
of Ref.~\protect\cite{DLMFL-98}, ``expt," and the small-$Q$ 
approximations, ``series3" and ``series4."
}
\label{fighQsmall}
\end{figure}

\begin{figure}[tb]
\hspace{-1cm}
\vspace{0cm}
\centerline{\psfig{width=10truecm,angle=-90,file=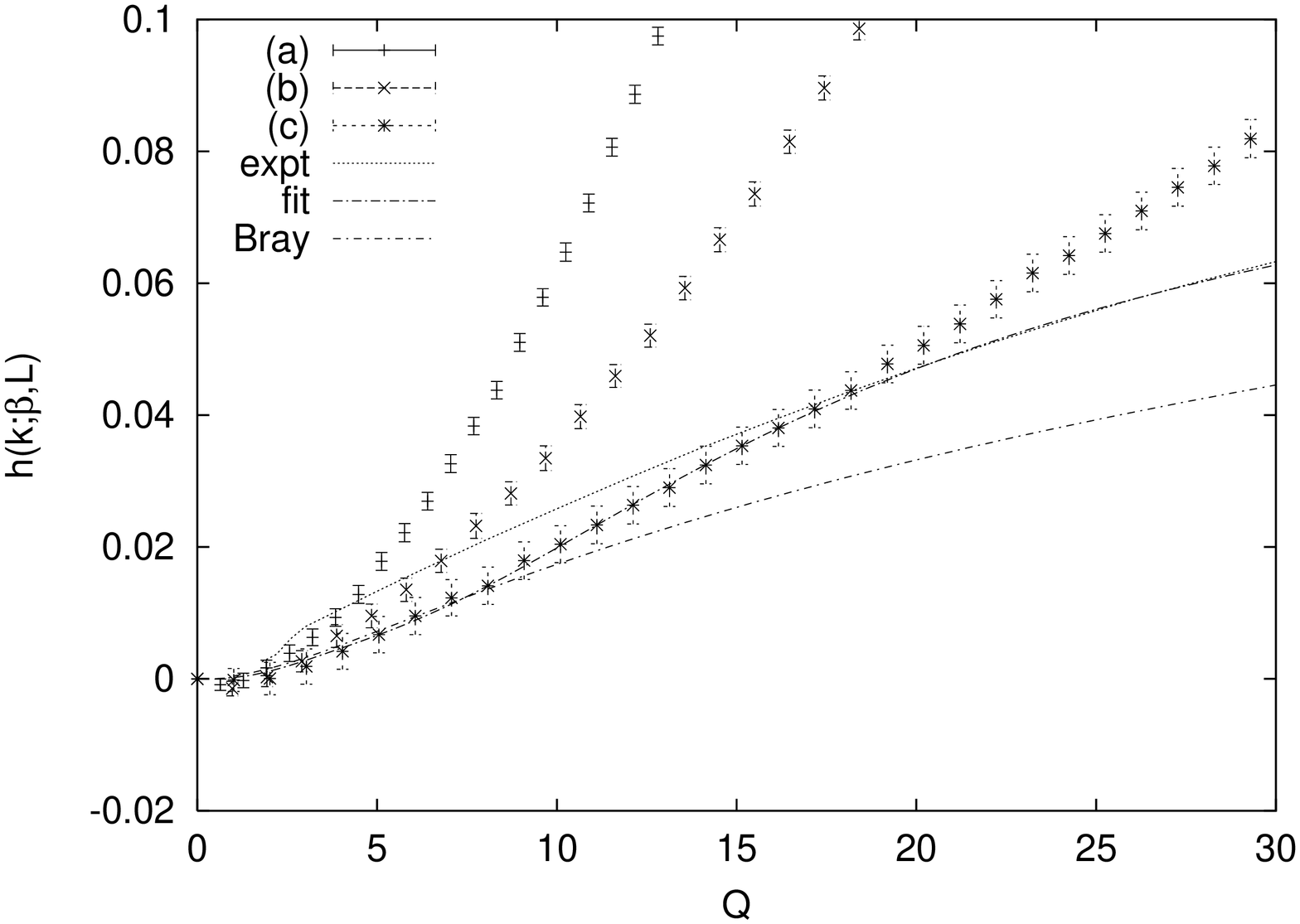}}
\vspace{0cm}
\caption{Function $h(q;\beta,L)$ versus $Q\equiv q \xi$ for the
three cases (a), (b), (c). We also report the experimental results
of Ref.~\protect\cite{DLMFL-98}, ``expt," a phenomenological 
interpolation , ``fit," and Bray's approximation, ``Bray."
}
\label{fighQ}
\end{figure}

However, at a closer look one observes tiny deviations of order 1-2\%. 
In order to observe better the differences among the different
approximations and data, it is useful to plot the function
$h(q;\beta,L)$ which converges to $\ln[(1 + Q^2) g_+(Q)]$ in the scaling limit.
We have been able to observe accurately (i.e. at the level of one 
error bar, approximately 0.3\% on $g_+(Q)$) this convergence only up to 
$Q\approx 4$, as it can be seen in Fig. \ref{fighQsmall}. 
Indeed, only in this region we observe a good overlap of the results 
for the two lattices (b) and (c), which have the largest values of $\xi$. 
As a further check, we can compare the numerical results with the 
small-$Q$ expansion (\ref{gQ-small}) which is expected to converge
rapidly up to $Q \approx 3$. Using Eq. (\ref{gQ-small}) to order $Q^6$ 
(resp. $Q^8$)
we obtain the curve labelled ``series3" (resp. ``series4") 
in Fig. \ref{fighQsmall}. 
The data (c), that correspond to $L=256$, are in perfect agreement, 
confirming that in this region we are seeing the correct asymptotic 
behavior.
In Fig. \ref{fighQsmall} we also report the experimental results of 
Ref. \cite{DLMFL-98}. They are systematically 
higher than the Monte Carlo results 
and indicate that, at least in this region,
the experimental error on the structure factor 
is approximately of order 0.5-1\%.

For larger values of $Q$, we are not able to observe scaling,
as it can be seen in Fig. \ref{fighQ}. 
According to standard renormalization-group theory
\begin{equation}
h(q;\beta,L) = h_1(Q,L/\xi) + L^{-\omega} h_2(Q,L/\xi) + \ldots
\end{equation}
where \cite{CPRV-02} $\omega = 0.83(5)$. Thus, we could try to extrapolate 
in $L$ at $L/\xi$ fixed and then take the limit $L/\xi\to \infty$. 
Lattices (b) and (c) have approximately the same $L/\xi$, 
$L/\xi\approx 6$ and thus, in principle one should be 
able to extrapolate in $L$. In practice, corrections increase quickly
with $Q$, see Fig. \ref{fighQ}, and no reliable extrapolation can be done.
In any case, we believe we can still use the numerical data presented in 
Fig. \ref{fighQ} to conclude conservatively that, for $Q\lesssim 15$-20,
$h(q;\beta,L)$ for lattice (c) is a good approximation to the limiting 
function with an error at most of 0.02, i.e. that we can use our data (c)
to compute $g_+(Q)$ with a 2\% precision up to $Q\lesssim 15$-20.

In Fig. \ref{fighQ} we also report Bray's approximation. 
Such an approximation agrees nicely with the Monte Carlo results
(c) up to $Q\approx 10$ and, as expected, it is lower in the region 
$Q\gtrsim 10$ where we expect the results (c) to be higher than
the scaling limiting curve. Bray's function looks therefore
a reasonable approximation to the universal scaling function, although it is 
somewhat lower than the experimental data by 1-2\%.

For the computations of the next Section, it is important to have an estimate 
of the structure factor with a reasonable error bar. For this purpose, 
we have determined a second interpolation that is in better agreement with 
the experimental data. We will obtain an error by comparing the results 
obtained using this interpolation and Bray's approximation.
This interpolation may be obtained by considering expressions 
that agree with the numerical data for lattice (c) in the region 
$Q < Q_{\rm max} \approx 15$.
We shall use again the spectral representation (\ref{gQ-dispersive}),
since such an expression gives automatically the behavior 
(\ref{pattern-cip1-ci}) and ensures 
the correct small-$Q$ behavior. In order to obtain the correct large-$Q$
behavior, we use a generalization of the spectral function proposed by 
Bray, i.e.
\begin{equation}
F_{\rm fit}(u) = F_B(u) \left(1 - u^{-2}\right) 
         \left(1 + \sum_{n=2}^{n_{\rm max}} a_n u^{-n}\right).
\label{Ffit}
\end{equation}
Such an expression is purely phenomenological. The first term has been 
introduced to guarantee that $F_{\rm fit}(1) = 0$ as generally expected, 
while corrections of order $1/u$ have been avoided, since they would 
give rise to terms of order $1/Q^{2-\eta-1}$ for $Q\to \infty$ 
that are stronger than those appearing in the Fisher-Langer behavior 
(\ref{eq:FL}). In Eqs. (\ref{gQ-dispersive}) and (\ref{PQ-def})
we use the $\epsilon$-expansion estimates (\ref{FL-par-eps}) 
and the values of the exponents reported in Eq. (\ref{exponents}).
The constants $a_n$ are fixed by requiring $g_+^{-1}(0)=1$ and 
$g_+(Q)$ to fit the numerical data (c) up to $Q\le 15$. A good fit 
is obtained by taking $n_{\rm max} = 6$ and 
$a_2= -574.128$, $a_3 =  7588.59$, $a_4=  -29558.9$,
$a_5= 43740.7$, $a_6=  -21715.6$. The corresponding curve labelled ``fit" is 
reported in Fig. \ref{fighQ}. The results depend on $Q_{\rm max}$
used in the fit, and tend to give a lower curve if smaller values of 
$Q_{\rm max}$
are used. However, it is interesting to remark that, with the
choice $Q_{\rm max}=15$, the interpolation is in excellent agreement with 
the experimental data for all $Q>15$, see Fig. \ref{fighQ}.

Finally, it is interesting to remark that the Ornstein-Zernicke 
approximation differs at most 1\% from the correct expression for 
$Q\lesssim 5$, while for $Q\gtrsim5$ the Fisher-Langer formula can be 
applied, as already observed in many experimental works,
see, e.g., Refs.~\cite{LMBWH-97,BC-97-98,BRCB-00,BCB-01}.

\begin{figure}[tb]
\hspace{-1cm}
\vspace{0cm}
\centerline{\psfig{width=10truecm,angle=90,file=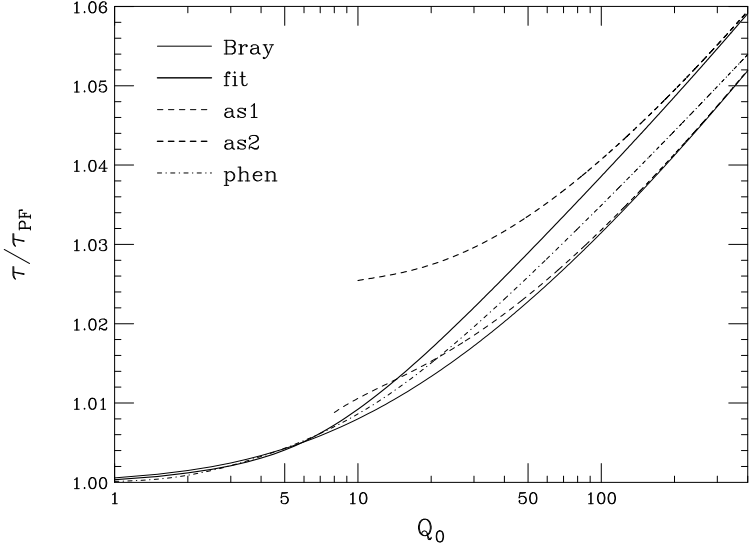}}
\vspace{0cm}
\caption{Ratio $\tau/\tau_{\rm PF}$ versus $Q_0$ using 
Bray's approximation, ``Bray," and the phenomenological approximation,
"fit."
We also report the corresponding asymptotic expression 
$\tau_{\rm as}/\tau_{\rm PF}$, (``as1" and ``as2")
where $\tau_{\rm as}$ is defined in Eq.~(\ref{turbidity-largeQ0}), 
and the phenomenological approximation (\ref{phen-turb}), ``phen,"
valid for $Q_0\le 100$.
In ``as1" we use $C_1^+ = 0.91797$, $K = 0.128735$, 
in ``as2" we use $C_1^+ = 0.92$, $K = 0.160734$.
}
\label{fig-turb}
\end{figure}

\section{Turbidity} \label{sec.4}

The turbidity $\tau$ is defined as the attenuation of the transmitted
light intensity per unit optical path length due to the scattering 
with the sample. Explicitly, it is given by
\begin{equation}
\tau \sim \int d\Omega\, S(k) \left[1 - {1\over2} \sin^2\theta\right],
\end{equation}
where $k = 2 k_0 \sin (\theta/2)$, $k_0 = 2 \pi n/\lambda$ is the 
momentum of the incoming radiation in the medium, $\lambda$ the 
corresponding wavelength in vacuum, $n$ the refractive index, and 
$\Omega = (\phi,\theta)$ the solid angle. By using Eq. (\ref{scaling-Sk}),
in the high-temperature phase
we can write the turbidity in the form
\begin{equation}
\tau = {2 \tau_0 t^{-\gamma}\over Q_0^2} \, 
   \int_0^{2 Q_0} QdQ\, g_+(Q) 
    \left[1 - {Q^2\over 2Q_0^2} + {Q^4\over 8 Q_0^4}\right],
\end{equation}
where $Q_0\equiv k_0 \xi$ and $\tau_0$ is a constant that can be 
assumed temperature-independent in a neighborhood of the critical point. 

For small values of $Q_0$, the Ornstein-Zernicke approximation can be used 
obtaining the Puglielli-Ford expression \cite{PF-70}
\begin{equation}
\tau_{\rm PF} = \tau_0 t^{-\gamma} 
  \left[ {2 a^2 + 2 a + 1\over a^3} \log(2 a + 1) - 
         {2 (a + 1)\over a^2}\right],
\end{equation}
where $a = 2 Q_0^2$.

We can also compute the behavior for large $Q_0$ by using Eq. (\ref{eq:FL}).
We obtain
\begin{equation}
\tau_{\rm as} = {2 \tau_0 t^{-\gamma}\over Q^2_0} 
  \left[C_1^+ (2 Q_0)^\eta 
    {\eta^2 + 2 \eta + 8\over \eta(\eta+2)(\eta+4)} - 
    {C_1^+\over \eta} + K + O\left(Q_0^{\eta - (1-\alpha)/\nu}\right)\right],
\label{turbidity-largeQ0}
\end{equation}
where 
\begin{equation}
K = \int_0^1 Q dQ g_+(Q) + 
    \int_1^\infty Q dQ\, \left[g_+(Q) - C_1^+ Q^{\eta-2}\right].
\end{equation}
In order to obtain $\tau$ for all values of $Q_0$ we must use a specific form 
for $g_+(Q)$. We will use here Bray's approximation and the 
interpolation formula obtained using (\ref{Ffit}) with $n_{\rm max} = 6$,
$Q_{\rm max} = 15$.
The difference between the results obtained using these two
expressions provides the error on our results. 
In Fig. \ref{fig-turb} we report $\tau/\tau_{\rm PF}$ using the 
two different approximations together with 
their asymptotic expression $\tau_{\rm as}/\tau_{\rm PF}$. 
In Bray's approximation $K = 0.128735$ while in the second one 
$K = 0.160734$.
The deviations from the Puglielli-Ford behavior
are very small and for $Q_0\gtrsim 100$ are well described by the 
asymptotic expression (\ref{turbidity-largeQ0}) with 
$C_1^+ \approx 0.92$ and $K = 0.145(16)$. 
Estimates of the turbidity for $1\lesssim Q_0 \lesssim 100$ can be found in 
Table \ref{table-turb}. For $Q \le 100$ one can use the phenomenological
formula
\begin{equation}
\tau = \tau_{\rm PF} \left[
   0.666421 + 0.242399 \left(1 + 0.0087936 Q_0^2\right)^{0.018195} + 
             0.0911801 \left(1 + 0.09 Q_0^4\right)^{0.0090975}\right],
\label{phen-turb}
\end{equation}
which is also reported in Fig. \ref{fig-turb} (``phen").

\begin{table}
\caption{Ratio $\tau/\tau_{\rm PF}$. We use here: (a) Bray's approximation;
(b) general phenomenological interpolation based on (\ref{Ffit})
with $n_{\rm max} = 6$ and $Q_{\rm max} = 15$.}
\label{table-turb}
\begin{tabular}{lcc}
$Q_0$ & (a)  & (b)  \\ 
\hline
  5 & 1.004 & 1.004 \\
 10 & 1.008 & 1.009 \\
 15 & 1.011 & 1.014 \\
 20 & 1.013 & 1.017 \\
 25 & 1.015 & 1.020 \\
 30 & 1.017 & 1.022 \\
 35 & 1.019 & 1.024 \\
 40 & 1.020 & 1.026 \\
 45 & 1.022 & 1.028 \\
 50 & 1.023 & 1.029 \\
 60 & 1.025 & 1.031 \\
 70 & 1.027 & 1.034 \\
 80 & 1.029 & 1.036 \\
 90 & 1.030 & 1.037 \\
100 & 1.032 & 1.039 \\
\end{tabular}
\end{table}

We wish finally to compare our results with the approximate expressions 
given by Ferrell \cite{Ferrell-91}, which require $Q_0 \gg 1$ and 
$\eta\log Q_0 \ll 1$, i.e. $1 \ll Q_0 \ll e^{1/\eta} \approx 9\times 10^{11}$.
By expanding Eq. (\ref{turbidity-largeQ0}) and setting as 
in Ref. \cite{Ferrell-91} $L = \log(4 Q_0^2)$ we obtain
\begin{equation}
\tau \approx {\tau_0 t^{-\gamma}\over Q_0^2} 
  \left[C_1^+ (L-1) + C_1^+ \eta \left({L^2\over4} - {L\over2} + 
   {3\over4}\right) + K\right].
\end{equation}
In order to compare with Ferrell's results, we must compute 
${\tau/(4\tau_0 t^{-\gamma} g(2 Q_0))}$. Since, using the same approximations
$g(2Q_0) = C_1^+(2 Q_0)^{-2} (1 + \eta L/2 + O(\eta^2) )$, we obtain
\begin{equation}
{\tau\over 4 \tau_0 t^{-\gamma} g(2 Q_0)} \approx
   L - 1 - {\eta L^2\over 4} + \eta \left({3\over4} + {K\over \eta C_1^+}\right)
\label{Ferrell}
\end{equation}
This formula agrees with Ferrell's expression once we recognize
that $K = O(\eta)$ since $K = 0$ for a purely Ornstein-Zernicke
behavior. Numerically, we predict $3/4 + K/(\eta C_1^+) \approx 5.1(5)$, 
which is smaller than Ferrell's numerical result 8.4. 
Ferrell's expression predicts a turbidity that is somewhat 
higher than ours.  Indeed, his numerical result implies 
$K \approx 0.26$ in Eq. (\ref{turbidity-largeQ0}), and as consequence we 
would obtain $\tau/\tau_{\rm PF}\approx 1.06$ (resp. 1.085) for $Q_0 = 100$ 
(resp. 1000), to be compared with our prediction
$\tau/\tau_{\rm PF}\approx 1.036(4)$ (resp. 1.069(3)).

Another expression for the turbidity that takes into 
account the anomalous decay of the structure factor is 
given in Ref. \cite{CLL-72}. It assumes that\cite{footCalmettes} 
$g_+(Q) = (1 + cQ^2)^{-1+\eta/2}$, where $c = 1/(1 - \eta/2)$. It follows
that 
\begin{equation}
\tau = 4 \tau_0 t^{-\gamma} 
{\left[(2 b + 1)^{\eta/2} - 1\right]\left[4 - 2 b(\eta-4) + 
         b^2 (\eta^2 + 2 \eta + 8)\right] - 4 \eta b(1+b) \over 
         b^3 \eta (2 + \eta) (4 + \eta)} ,
\end{equation}
where $b = 4 Q_0^2/(2 - \eta)$. Such an expression however predicts a 
turbidity that is too large. For instance, for $Q_0 = 10$ it gives 
$\tau /\tau_{\rm PF} \approx 1.05$, to be compared with our prediction
$\tau /\tau_{\rm PF} \approx 1.008$, cf. Table \ref{table-turb}.

Note the correct turbidity $\tau$ is larger than $\tau_{\rm PF}$
since $g_+(Q)$ decreases slower for $Q\to \infty$ than the 
Ornstein-Zernicke approximation. However, this is apparently in contrast
with the experimental results for the binary fluid mixture 
methanol-cyclohexane presented in Ref. \cite{JLMW-99}.

\section*{Acknowledgments}

The work of V.M.M. was supported by the European Commission contract
HPMF-CT-2000-00450 and by OCYT (Spain), project No. FPA2001-1813.


\begin{references}
\bibitem{foot1}
The structure factor directly determines the scattered light intensity 
in the absence of multiple scattering. However, multiple scattering is 
important in experiments on fluids \cite{SS-88,Anisimov-book}, 
and it is therefore essential to perform the appropriate corrections 
to the data \cite{SS-88,Anisimov-book,BC-94,Cipelletti-97}.
Note also that we consider here only Rayleigh elastic scattering, 
which is the dominant contribution near the critical point, see, e.g.,
Ref.~\cite{Anisimov-book}.
\bibitem{footCalmettes}
In Ref. \cite{CLL-72}, the scaling function $g_+(Q)$ is written 
as $g_+(Q) = 1/(1 + q^2 \xi^2)^{1-\eta/2}$, so that the 
correlation length defined there does not coincide with the 
usual second-moment one. In order to be consistent with 
our normalizations, we have introduced the correction factor $c$.
\end{references}
\end{document}